\documentclass[doublecol]{epl2} 

\title{Geometric quantum phase for displaced states for a particle with an induced electric dipole moment}
\shorttitle{Geometric quantum phase for displaced states } 

\author{J. Lemos de Melo\inst{1,2} \and K. Bakke\inst{2} \and C. Furtado\inst{2}}
\shortauthor{ J. Lemos de Melo \etal}

\institute{                    
  \inst{1} Centro de Ci\^encias Exatas e Tecnol\' ogicas, Universidade Federal do Reconcavo da Bahia,
Cruz das Almas, BA, Brazil\\
  \inst{2} Departamento de F\'isica, Universidade Federal da Para\'iba, Caixa Postal 5008, 58051-970, Jo\~ao Pessoa, PB, Brazil.
}

\abstract{Basing on the analogue  Landau levels  for a neutral particle possessing an induced electric dipole moment,  we show  that displaced states can be built in the presence of electric and magnetic fields. Besides, the Berry phase associated with these displaced quantum states is obtained by performing an adiabatic cyclic evolution in series of paths in parameter space. }

\begin{document}

\maketitle

In 1984, Berry \cite{berry} investigated the evolution of a quantum state of a system during an adiabatic cyclic evolution and showed that this quantum state acquires a phase shift when the system returns to its initial state. Now, we consider that the Hamiltonian operator  depends on a set of parameters.  In short,   if the  these parameters  slowly vary  during the evolution of a quantum system  and then return to their initial  values, the quantum system returns to its initial state up to a phase. This phase shift depends only on the geometrical nature of the path  of evolution of the quantum system, therefore, it became known as geometric quantum phase or Berry phase. Moreover, Berry \cite{berry} showed that the quantum phase  giving rise to the Aharonov-Bohm effect \cite{ab} is a special case of the geometric quantum phase. Indeed, the Aharonov-Bohm phase \cite{ab} is a topological quantum phase because it is independent of the path taken in the parameter space. On the other hand, the study of geometric quantum phases was extended to non-adiabatic cyclic evolutions by Aharonov and Anandan \cite{ahan} in 1987. Wilczek and Zee \cite{wilk}, in turn, generalized the Berry phase for non-Abelian cases.  To illustrate this geometric phase, we consider  a  subspace spanned  by the eigenvectors of a family of Hamiltonians $\mathcal{F}=\left\{H\left(\lambda\right)=U\left(\lambda\right)H_{0}U^{\dag}\left(\lambda\right);\lambda\in
\mathcal{M}\right\}$, where $U\left(\lambda\right)$ is the unitary operator and $\lambda$ corresponds to the control parameter \cite{zr1,pzr1,pc1,pc3}.  Now we obtain the geometric phase performing the following procedure:  we suggest that  the control parameters are changed  adiabatically along a loop in the control manifold $\mathcal{M}$. The action of the unitary operator $U\left(\lambda\right)$ on an initial state $\left|\psi_{0}\right\rangle$ brings it to a final state $\left|\psi\right\rangle=U\left(\lambda\right)\left|\psi_{0}\right\rangle$ . The general expression  corresponding to the action of this unitary operator is  $ \left|\psi\right\rangle=U\left(\lambda\right)\left|\psi_{0}\right\rangle=e^{-i\int_{0}^{t}E\left(t'\right)\,dt'}\,\Gamma_{A}\left(\lambda\right)\left|\psi_{0}\right\rangle,
\label{eq:}$ where the first factor corresponds to the dynamical phase, and the second factor is the holonomy: 
$
\Gamma_{A}\left(\lambda\right):=\hat{\mathcal{P}}\,\exp i \int_{C} \mathcal{A},$
 where $\mathcal{A}=A\left(\lambda\right)\,d\lambda$ is a connection $1$-form called the Mead-Berry connection 1-form \cite{tg1}. Notice that the expression $\Gamma_{A}$ is a geometric phase well-known as Berry's Phase. This phase generally depends on paths $C$ and depends  on the geometrical nature of the pathway along which the system evolves. The quantity $A\left(\lambda\right)$ corresponds to the Mead-Berry vector potential, where the components of this vector potential are defined as: $A^{\alpha\beta}=i \left\langle \psi^{\alpha}\left(\lambda\right)\right|\partial/\partial\lambda\left|\psi^{\beta}\left(\lambda\right)\right\rangle$. Besides, the  observations of adiabatic and non-adiabatic Berry quantum phases have been reported in several areas of physics such as photons systems \cite{16,16a}, neutrons \cite{17} and nuclear spins \cite{18}. Several authors have investigated, from a theoretical viewpoint, the manifestation of Berry phases in various areas of physics \cite{19,20,bf21,epjcj}. Recently, Yang and Chen \cite{displaced1} have obtained the Abelian and non-Abelian Berry phases associated with displaced Fock\cite{displaced} states for Landau levels.

Recently,  essential efforts have been devoted to investigating the behaviour of neutral particle systems in a situation analogous to the Landau quantization \cite{landau}. The first system was proposed by Ericsson and Sj\"oqvist \cite{er} by dealing with a neutral particle with a permanent magnetic dipole moment  exposed to an electric field. The interaction between the magnetic dipole moment of the neutral particle and the electric field is described by the Aharonov-Casher coupling \cite{ac} and became known as the Landau-Aharonov-Casher quantization. By following this line of research, Ribeiro {\it et al.} \cite{lin} considered a neutral particle with a permanent electric dipole moment and,  being based on the He-McKellar-Wilkens effect \cite{HMW,HMW2}, proposed the Landau-He-McKellar-Wilkens quantization. Moreover, the Landau quantization for a neutral particle  possessing an induced electric dipole moment was investigated by Furtado {\it et al.} \cite{lwhw}. It is worth mentioning that the Landau quantization for a neutral particle possessing an induced electric dipole moment was based on the system proposed by Wei {\it et al.} \cite{wei,c1lwhw} in a study of geometric quantum phases. Recently, an experimental test of the He-McKellar-Wilkens effect \cite{HMW,HMW2} has been reported by Lepoutre {\it et al.} \cite{lepoutre,lepoutre1,lepoutre2,lepoutre3}, where the field configuration proposed by Wei {\it et al.} \cite{wei} was used in the experiment.

In this  paper, we consider a neutral particle  with an induced electric dipole moment  interacting with electric and magnetic fields. We build,  for the first time in the literature, the  displaced Fock states from the Landau states obtained in Ref. \cite{lwhw}  through the approach introduced by Feldman and Kahn \cite{feld}. Then, we investigate the arising of Abelian and non-Abelian geometric quantum phases associated with these displaced Fock states during some adiabatic cyclic evolutions. It is worth  to mention that the field configuration giving rise to the Landau quantization of neutral particle with an induced electric dipole moment provides a  practical  perspective where geometric quantum phases can be used to investigate a way of performing the holonomic quantum computation \cite{zr1} in atomic systems. We emphasize that  the displaced states and  the Abelian and non-Abelian phases are new  results of this present  contribution.
This paper is organized as follows. In section II, we make a brief review of the Landau quantization for a neutral particle possessing an induced electric dipole moment. In section III, we build the displaced Fock states based on the Landau system for a neutral particle with an induced electric dipole moment. In section IV, we study the arising of Abelian and non-Abelian geometric phases associated with the displaced Fock states.  Finally, in section V, we present our conclusions.
\section{Landau quantization}
In this section, we make a brief review of the Landau quantization for a neutral particle possessing an induced electric dipole moment. In \cite{wei} the quantum dynamics of a moving particle with an induced electric dipole moment is described as follows: first, let us suggest the dipole moment of an atom to be proportional to the external electric field in the rest frame of the particle or in the laboratory frame, that is, $\vec{d}=\alpha\,\vec{E}$, where $\alpha$ is the dielectric polarizability of the atom; thus, for a moving particle, the dipole moment of the neutral particle interacts with a different electric field $\vec{E}'$, which is given by applying the Lorentz transformation of the electromagnetic field: $\vec{E}'=\vec{E}+\frac{1}{c}\,\vec{v}\times\vec{B}$ up to $\mathcal{O}\left(v^{2}/c^{2}\right)$, where $\vec{v}$ is the velocity of the neutral particle and the fields $\vec{E}$ and $\vec{B}$ correspond to the electric and magnetic fields in the laboratory frame, respectively. From now on, let us consider the units $\hbar=c=1$, therefore, we can write $
\vec{d}=\alpha\left(\vec{E}+\vec{v}\times\vec{B}\right).\label{1.1}$ Further, the Lagrangian is given by $\mathcal{L}=\frac{1}{2}\left(M+\alpha\,B^{2}\right)\,v^{2}+\vec{v}\cdot\left(\vec{B}\times\alpha\,\vec{E}\right)+\frac{1}{2}\,\alpha\,E^{2}$,  {where we suggested  that velocity of the dipole} is perpendicular  to the magnetic field. Note that this moving particle has an effective mass given by $m=M+\alpha\,B^{2}$, where $M$ is the mass of the neutral particle. Let us consider $B^{2}=B_{0}^{2}=\mathrm{constant}$. Thereby, after some calculations, the Schr\"odinger equation  describing the quantum dynamics of a moving neutral particle with an induced electric dipole moment interacting with electric and magnetic fields can be written in the form:
\begin{eqnarray}
i\frac{\partial\psi}{\partial t}=\frac{1}{2m}\left[\vec{p}+\alpha\left(\vec{E}\times\vec{B}\right)\right]^{2}\psi-\frac{\alpha}{2}\,E^{2}\,\psi.
\label{1.2}
\end{eqnarray}
Now, basing on the discussions  presented in Refs. \cite{wei,c1lwhw} and  considering particles with a large polarizability $\alpha\sim10^{-39}\,\mathrm{Fm}^{2} $ and $B\leq10\,\mathrm{T} $, we find that the term $\alpha B^{2}\leq 10^{-37}\,\mathrm{kg}$ corresponds to only $ 10^{-10}$ of the mass of one nucleon, therefore this term has no significance in the  Hamiltonian operator of the right-hand-side of Eq. (\ref{1.2}). Moreover, by considering an electric field  with intensity $E\approx10^{7}\,\mathrm{V}/\mathrm{m}$,  one sees that the term $ \alpha E^{2}=10^{-25}\,\mathrm{J}$ is very small  being compared with the kinetic energy of the atoms, hence, it can be neglected in the Hamiltonian given in Eq. (\ref{1.2}) without loss of generality. In this way, we can write the Hamiltonian of Eq. (\ref{1.2}) in the form:
\begin{eqnarray}
\hat{\mathcal{H}}=\frac{1}{2M}\left[\vec{p}+\alpha\left(\vec{E}\times\vec{B}\right)\right]^{2}.
\label{1.3}
\end{eqnarray}
The Landau quantization, in  its turn, was established in Ref. \cite{lwhw}  to exhibit  the well-known Landau levels for systems of cold atoms \cite{cold1,cold2,cold3,cold4,cold5}. In this case, a cold atom is treated as a structureless particle with an induced electric dipole moment. The field configuration proposed in Ref. \cite{lwhw} in order to achieve the Landau quantization for a neutral particle with an induced electric dipole moment is 
$
\vec{E}=\frac{\lambda}{2}\left(x,y,0\right)\quad\mbox{and}\quad\vec{B}=\left(0,0,B\right),
\label{EeB}
$
where $\lambda$ and $B$ are the uniform electric charge density and the magnetic field intensity, respectively. With this field configuration, we have an effective vector potential defined as $\vec{A}_\mathrm{eff}=\vec{E}\times\vec{B}=\frac{\lambda\,B}{2}\left(y,\,-x,\,0\right)$ and, thus, an effective magnetic field given by $\vec{B}_\mathrm{eff}=-\lambda\,B\,\left(0,0,1\right)$. Note that, with the field configuration (\ref{EeB}), the interaction between the electromagnetic field and the electric dipole moment of the atom described by the Hamiltonian operator (\ref{1.3}) is analogous to the minimal coupling of a charged particle  to an external magnetic field. Hence, this field configuration yields the precise condition in which the Landau quantization occurs in a cold atom system, since the quantum particle is constrained to move in a plane in the presence of an effective uniform magnetic field $\vec{B}_\mathrm{eff}=-\lambda\,B\,\left(0,0,1\right)$. It is worth mentioning recent studies that have dealt with effective vector potentials and effective magnetic fields in order to obtain the Landau-Aharonov-Casher quantization \cite{er,bf20} and the Landau-He-McKellar-Wilkens quantization \cite{lin}. Thereby, by using the field configuration given in Eq. (\ref{EeB}), the Hamiltonian operator (\ref{1.3}) can be written in the form:
\begin{eqnarray}
\hat{\mathcal{H}}=\frac{1}{2M}\left[\left(p_{x}+\frac{M\omega}{2}\,y\right)^{2}+\left(p_{y}-\frac{M\omega}{2}\,x\right)^{2}\right],
\label{Hcoml2}
\end{eqnarray}
where $\omega=\omega_{\mathrm{WHW}}=\frac{\alpha\,\lambda\,B} {M}=\sigma\,\left|\omega\right|$ is the cyclotron frequency of the corresponding Landau levels for a neutral particle with an induced electric dipole moment \cite{lwhw} and $\sigma=\mathrm{sgn}(\lambda\,B)$ indicates the direction of  rotation. 

\section{Displaced Fock states}
The concept of the coherent states was proposed originally in 1926 by Schr\"odinger \cite{schro} in the  context of  classical states of the quantum  harmonic oscillator.  Klauder \cite{klauder1,klauder2}, Glauber \cite{glauber} and Sudarshan \cite{sudar},  independently have developed the concepts of coherent states  in quantum mechanics. Klauder and Sudarshan \cite{klauder} have  discussed how coherent states  can be constructed from the Fock vacuum through the action of the displacement operators.
On the other hand, Venkata Satyanarayana \cite{ven}  described states of the harmonic oscillator through the action of the displacement operator on the Fock states, which are called as displaced Fock states. In this letter we use this denomination for displaced Fock states that are constructed from the Fock state $n$, in order to differ them from coherent states constructed  from the Fock vacuum. The displaced Fock states have attracted a great interest in several physical systems \cite{2,3,4,5,6,babi,homodyne,banas,kuzm}. In particular, the displaced Fock states have been used  within studies of geometric quantum phases \cite{displaced} and holonomic quantum computation \cite{lbf}. In this section,  we concentrate  on the construction of displaced Fock states for an analogue of the Landau quantization for a neutral particle with an induced electric dipole moment. We follow the formalism adopted in Ref. \cite{feld} in order to build these states in the presence of an electric and a magnetic fields. First of all, let us introduce the following operators:
\begin{eqnarray}
a_{\pm}&=&\frac{l_m}{\sqrt{2}\hbar}\left[p_{x}\mp i\sigma p_{y}\pm \frac{i\hbar}{2l_{m}^{2}}\left(x\mp i\sigma\,y\right)\right],\nonumber\\
[-2mm]\label{b1}\\[-2mm]
b_{\pm}&=&\frac{1}{\sqrt{2}\,l_{m}}\left[\frac{1}{2}\left(x\mp i\sigma\,y\right)\pm \frac{il_{m}^{2}}{\hbar}\left(p_{x}\mp i\sigma\,p_{y}\right)\right],\nonumber
\end{eqnarray}
where we have defined the parameter $l_{m}=\sqrt{\hbar/M\,\left|\omega\right|}$ as the magnetic length and $\sigma=\pm1$. In particular, the operators $b_{\pm}$ are built from the orbit
center-coordinate operators $\hat{X}=\frac{1}{\sqrt{2}\,l_{m}}\left(\frac{1}{2}\hat{x} +\frac{l_m{}^2}{\hbar}\hat{p_{y}}\right)$ and $\hat{Y}=\frac{1}{\sqrt{2}\,l_{m}}\left(\frac{1}{2}\hat{y}-\frac{l_m{}^2}{\hbar}\hat{p_{x}}\right)$ via relations: $b_{\pm}=\hat{X}\mp i\,\sigma\,\hat{Y}$. Moreover, the operators $a_{\pm}$ and $ b_{\pm}$ obey the following relation of commutation:
\begin{eqnarray}
\left[a_{i}\,,\,b_{i}\right]=\left[a_{i}\,,\,b_{j}\right]=0;\,\,\,\,\,\,\left[a_{-}\,,\,a_{+}\right]=\left[b_{+}\,,\,b_{-}\right]=1.
\end{eqnarray}
Thereby, from Eq. (\ref{b1}), the Hamiltonian operator (\ref{Hcoml2}) and the $z$-component of angular momenta $\hat{L}_{z}$ can be rewritten as:
\begin{eqnarray}
\hat{\mathcal{H}}=\hbar\,\left|\omega\right|\,\left(a_{+}\,a_{-}+\frac{1}{2}\right);\,\,\,\,\,\,\hat{L}_{z}=\sigma\,\hbar\left(b_{-}b_{+}-a_{+}a_{-}\right),
\label{HeLcomab}
\end{eqnarray}
where we have that $\left[\hat{\mathcal{H}},\,\hat{L}_{z}\right]=0$. Since $\hat{\mathcal{H}}$ commutes with $\hat{L}_{z}$, then, these operators share a set of eigenstates; thus,
\begin{eqnarray}
\hat{\mathcal{H}}\,\left|n,\,\ell\right\rangle=\mathcal{E}_{n}\,\left|n,\,\ell\right\rangle;\,\,\,\,\hat{L}_{z}\,\left|n,\,\ell\right\rangle=\ell\,\hbar\,\left|n,\ell\right\rangle,
\label{HeLcomEel}
\end{eqnarray} 
where $\mathcal{E}_{n}=\hbar\,\left|\omega\right|\,\left(n+\frac{1}{2}\right)$, $n=0,\,1,\,2,\,\ldots$ and $\ell=0,\,\pm1,\pm2,\,\pm3,\ldots$. Besides, we have 
\begin{eqnarray}
\left[\hat{\mathcal{H}},\,a_{\pm}\right]=\pm\hbar\,\left|\omega\right|a_{\pm};\,\,\,\,\left[\hat{L}_{z},\,a_{\pm}\right]=\mp\,\sigma\,\hbar\,a_{\pm},
\label{ComHeLcoma}
\end{eqnarray}
which means, from the first commutation relation given in Eq. (\ref{ComHeLcoma}), that the operators $a_{+}$ and $a_{-}$ correspond to the raising and lowering operators  for energy, respectively. However, from the second commutation relation given in Eq. (\ref{ComHeLcoma}), by taking $\sigma=-1$,  it follows that the operators $a_{+}$ and $a_{-}$ play the role of the raising and lowering operators with respect to the eigenstates for $\hat{L}_{z} $, respectively; on the other hand, by taking $\sigma=+1$, the behaviour of these operators is inverted. Thereby, let us write
\begin{eqnarray}
a_{+}\,\left|n,\,\ell\right\rangle&=&\sqrt{n+1}\,\left|n+1,\,\ell-\sigma\right\rangle\nonumber \\,a_{-}\,\left|n,\,\ell\right\rangle&=&\sqrt{n}\,\left|n-1,\,\ell+\sigma\right\rangle.
\label{anl}
\end{eqnarray}

We can also observe that $\left[\hat{\mathcal{H}},\,b_{\pm}\right]=0$ and $\left[\hat{L}_{z},\,b_{\pm}\right]=\mp\,\sigma\,\hbar\,b_{\pm}$; thus, the operators $b_{\pm}$ can raise or lower only the eigenstates of $\hat{L}_{z}$ in such a way that
\begin{eqnarray}
b_{+}\,\left|n,\,\ell\right\rangle&=&\sqrt{n+\sigma\,\ell}\,\left|n,\,\ell-\sigma\right\rangle\nonumber\\
b_{-}\,\left|n,\,\ell\right\rangle&=&\sqrt{n+\sigma\ell+1}\,\left|n,\,\ell+\sigma\right\rangle,
\label{bnl}
\end{eqnarray}
which  shows that the operators $b_{+}$ and $b_{-}$ play the role of the raising and lowering operators with respect to the eigenstates of $\hat{L}_{z}$, respectively, for $\sigma=-1$. The behaviour of these operators is inverted for $\sigma=+1$. 

From the relations established in Eq. (\ref{bnl}), by considering $\sigma=-1$,  one can use the first relation of Eq. (\ref{bnl})  to show  that for $\ell=n$ we have that $b_{+}\left|n,\,n\right\rangle=0$ \cite{feld}. This means that the possible values of the quantum number $\ell$ are defined in the range $-\infty\,<\,\ell\,\leq\,n$. On the other hand, by considering $\sigma=+1$ and $\ell=n$, we obtain the range $-n\,\leq\,\ell\,<\,+\infty$. Hence, we can see that the sum $n+\sigma\,\ell$ is always a positive number. Basing on the algebraic method, one can define the ground state $\left|0,0\right\rangle$ in such a way that $a_{-}\left|0,0\right\rangle=b_{-}\left|0,0\right\rangle=0$, therefore we can write
\begin{eqnarray}
\left|n,\,\ell\right\rangle=\frac{a_{+}^{n}b_{+}^{n+\sigma\,\ell}}{\sqrt{n!\,\left(n+\sigma\ell\right)!}}\,\left|0,\,0\right\rangle,
\label{nlstate}
\end{eqnarray}
and the corresponding the ground state given in the coordinate representation  look like
\begin{eqnarray}
\psi_{00}=\left\langle x,y\right|\left.0,0\right\rangle=\frac{1}{2\,l_{m}^{2}\,\sqrt{\pi}}\,\exp\left[-\frac{\left(x^{2}+y^{2}\right)}{4\,l_{m}^{2}}\right].
\label{psi00}
\end{eqnarray}

From now on, let us build the displaced states for the Landau system of a neutral particle possessing an induced electric dipole moment.  Following Refs. \cite{displaced,displaced1}, the displaced states can be  constructed by applying the following unitary operator:
\begin{eqnarray}
\hat{D}\left(\nu\right)&=&\exp\left(\nu\,a_{+}-\nu^{*}\,a_{-}\right)\nonumber\\
[-2mm]\label{Dnu}\\[-2mm]
&=& e^{-\left|\nu\right|^{2}/2}\,e^{\nu\,a_{+}}\,e^{-\nu^{*}\,a_{-}},\nonumber
\end{eqnarray}
where $\nu=\nu_{x}+i\,\nu_{y}$, and the displaced states are given in the form: $\left|n\left(\nu\right),\,\ell\right\rangle=\hat{D}\left(\nu\right)\,\left|n,\,\ell\right\rangle$. In this way, the time evolution of the displaced states of the Landau system of a neutral particle possessing an induced electric dipole moment is governed by the Hamiltonian operator given by
\begin{eqnarray}
\hat{\mathcal{H}}_{\nu}&=&\hat{D}\left(\nu\right)\,\hat{\mathcal{H}}\,\hat{D}^{\dagger}\left(\nu\right)\nonumber\\
&=&\hbar\,\left|\omega\right|\left[\left(a_{+}-\nu^{*}\right)\left(a_{-}-\nu\right)+\frac{1}{2}\right]\label{Hnu}\\
&=&\frac{1}{2m}\left[\left(\hat{p}_{x}+\frac{\hbar}{2l_{m}^{2}}\,\hat{y}-\frac{\sqrt{2}\,\hbar}{l_{m}}\,\nu_{x}\right)^{2}\right. \nonumber \\ &+&\left. \left(\hat{p}_{y}-\frac{\hbar}{2l_{m}^{2}}\,\hat{x}+\frac{\sqrt{2}\,\hbar}{l_{m}}\,\nu_{y}\right)^{2}\right].\nonumber
\end{eqnarray}

Note that  two new terms in the Hamiltonian given by Eq. (\ref{Hnu}) can be treated as new contributions to the analogue of the vector potential $\vec{A}_{\mathrm{eff}}=\vec{E}\times\vec{B}$ given in Eq. (\ref{1.3}). From the experimental viewpoint, new contributions to this analogue of the vector potential can be achieved by adding a constant electric field parallel to the plane of motion of the neutral particle. This is reasonable because the conditions established in Ref. \cite{lwhw} for achieving the Landau quantization for a neutral particle with an induced electric dipole moment continue to be satisfied. Hence, by including a constant electric field $\vec{E}'=\left(E_{x}',E_{y}',0\right)$, one finds the parameters $\nu_{x}$ and $\nu_{y}$ given in (\ref{Hnu})  to be
\begin{eqnarray}
\nu_{x}=-\frac{\alpha\,l_{m}}{\sqrt{2}\,\hbar}\,E_{y}';\,\,\,\,\nu_{y}=-\frac{\alpha\,l_{m}}{\sqrt{2}\,\hbar}\,E_{x}'.
\label{nuxy}
\end{eqnarray}

Observe that the modification of the Hamiltonian given in Eq. (\ref{Hnu})  related to the parameters defined in Eq. (\ref{nuxy}) corresponds to the transformation $\left(x,y\right)\rightarrow\left(x+\delta x,\,y+\delta y\right)$ in the wave function, where $\delta x=\frac{2\,\alpha\,l_{m}^{2}}{\hbar}\,E_{x}'$ and $\delta y=\frac{2\,\alpha\,l_{m}^{2}}{\hbar}\,E_{y}'$. Hence, the electric field $\vec{E}'$ causes a small shift in the states of the Landau quantization of a neutral particle with an induced electric dipole moment in the phase space in a  way analogous to the Landau states discussed in Ref. \cite{displaced1}, where we must assume that this shift  takes place in a real space.

\section{Berry phase}

\begin{figure}[!h]
\includegraphics[scale=0.4]{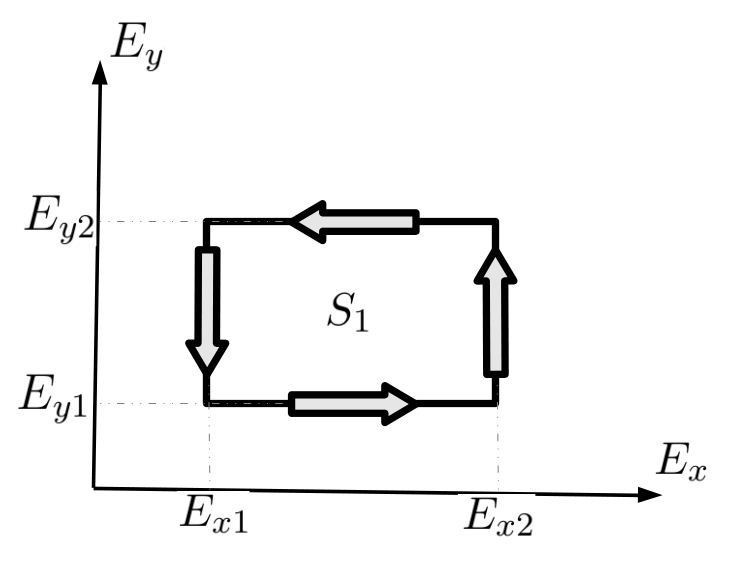}
\caption{Closed path of the adiabatic evolution of the control parameters $E_{x}'$ and $E_{y}'$ in the $E_{x}'-E_{y}'$ plane.}
\label{fig1}
\end{figure}

 Now let us study the Berry phase associated with the displaced states for the Landau system of a neutral particle possessing an induced electric dipole moment.  It follows from \cite{wilk,tg1} that due to an adiabatic cyclic evolution in a degenerated space, the wave function of a quantum particle acquires a non-Abelian geometric phase. From the adiabatic theorem, the connection 1-form is given by 
\begin{eqnarray}
A_{n}^{k,\ell}\left(\xi\right)=i \left\langle n\left(\nu\right),k \left|\frac{\partial}{\partial\xi}\right| n\left(\nu\right),\ell\right\rangle,
\label{AdeBerry}
\end{eqnarray}
where the parameter $\xi$ is the control parameter of the system. In the present case, the control parameters are defined by the components $E_{x}'$ and $E_{y}'$ of the electric field $\vec{E}'$, the electric charge density $\lambda $ and the magnetic field intensity $B$ established in Eq. (\ref{EeB}). Let us simplify our discussion by assuming that the control parameters are positive. Thereby, the non-{\bf zero} components of the connection 1-form (\ref{AdeBerry}) for an energy level $n$ are determined by
\begin{eqnarray}
&&A_{n}^{k,\ell}\left(\xi\right)=-\left(\nu_{x}\frac{\partial\nu_{y}}{\partial\xi}-\nu_{y}\frac{\partial\nu_{x}}{\partial\xi}\right)\delta_{k,\ell}\\
&-&\frac{1}{l_{m}}\frac{\partial\,l_{m}}{\partial\xi}\left(\nu^{*}\sqrt{n+\sigma\ell+1}\,\,\,\delta_{k,\ell+\sigma}+\nu\sqrt{n+\sigma\ell}\,\,\,\delta_{k,\ell-\sigma}\right).\nonumber
\label{AdeXi}
\end{eqnarray}
By handling the control parameters $E_{x}'$, $E_{y}'$, $\lambda$ and $B$, we obtain
\begin{eqnarray}
A_{n}^{k,\ell}\left(E_{x}'\right)&=&-\frac{E_{y}'}{16\,u^{2}\,\lambda\,B}\,\delta_{k,\ell};\nonumber\\
A_{n}^{k,\ell}\left(E_{y}'\right)&=&-\frac{E_{x}'}{16\,u^{2}\,\lambda\,B}\,\delta_{k,\ell};\nonumber\\
[-2mm]\label{AdeE}\\[-2mm]
A_{n}^{k,\ell}\left(\lambda\right)&=&-\frac{u}{\lambda^{3/2}\,B^{1/2}}\left[\left(E_{y}'-iE_{x}'\right)\sqrt{n+\sigma\ell+1}\,\delta_{k,\ell+\sigma}\right. \nonumber\\&+&\left.\left(E_{y}'+iE_{x}'\right)\sqrt{n+\sigma\ell}\,\delta_{k,\ell-\sigma}\right];\nonumber\\
A_{n}^{k,\ell}\left(B\right)&=&-\frac{u}{\lambda^{1/2}\,B^{3/2}}\left[\left(E_{y}'-iE_{x}'\right)\sqrt{n+\sigma\ell+1}\,\delta_{k,\ell+\sigma}\right. \nonumber \\ &+& \left.\left(E_{y}'+iE_{x}'\right)\sqrt{n+\sigma\ell}\,\delta_{k,\ell-\sigma}\right],\nonumber
\end{eqnarray}
where $u=\sqrt{\frac{\hbar}{8\alpha}}$. In particular, the connections 1-forms $A_{n}^{k,\ell}\left(E_{x}'\right)$ and $A_{n}^{k,\ell}\left(E_{y}'\right)$ given in Eq. (\ref{AdeE}) are  called the Berry connections or the Mead-Berry vector potentials \cite{tg1,berry}, and these  connection  give rise to Abelian geometric quantum phases. In this way, by taking $k=\ell$ and keeping the control parameters $\lambda$ and $B$ unchanged,  one finds the  corresponding quantum phase $\Gamma(c_{1})= e^{i\gamma_{n}}$, where phase angle $\gamma_{n}$ is given by
\begin{eqnarray}
\gamma_{n}^{\left(1\right)}&=&\oint_{C_{1}}\left[A_{n}^{k,\ell}\left(E_{x}'\right)\,dE_{x}'+A_{n}^{k,\ell}\left(E_{y}'\right)\,dE_{y}'\right]\nonumber\\
&=&-\frac{1}{16\,u^{2}\,\lambda\,B}\,S_{1},
\label{gamma1}
\end{eqnarray}
where $C_{1}$ is the path of the adiabatic cyclic evolution taken in the $E_{x}'-E_{y}' $ plane and $S_{1}$ is the area enclosed by the path $C_{1}$ as shown in Fig. \ref{fig1}. Note that the expression (\ref{gamma1}) produces an Abelian geometric phase.

\begin{figure}[!h]
\includegraphics[scale=0.3]{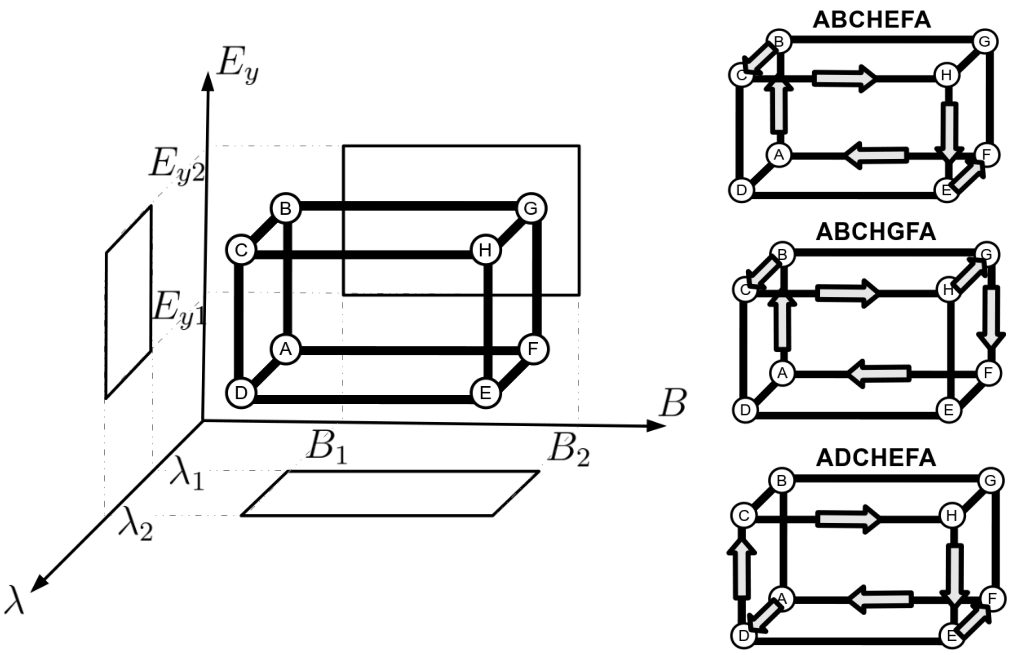}
\caption{Possible paths of the adiabatic evolution in the three parameters space.}
\label{fig2}
\end{figure}

In what follows, let us deal with the non-diagonal terms of the connection 1-form defined in Eq. (\ref{AdeXi}). The non-diagonal terms are defined by $A_{n}^{k,\ell}\left(\lambda\right)$ and $A_{n}^{k,\ell}\left(B\right)$ given in Eq. (\ref{AdeE}) by taking $k\neq\ell$. A particular case is given by considering $E_{x}'=0$; thus, the corresponding phase angle is
\begin{eqnarray}
\gamma_{n}^{\left(i\right)}&=&\oint_{C_{2}}\left[A_{n}^{k,\ell}\left(E_{y}'\right)\,dE_{y}'+A_{n}^{k,\ell}\left(\lambda\right)\,d\lambda+A_{n}^{k,\ell}\left(B\right)\,dB\right]\nonumber\\
&=&2u\left[\sqrt{n+\sigma\ell+1}\,\,\delta_{k,\ell+\sigma}+\sqrt{n+\sigma\ell}\,\,\delta_{k,\ell-\sigma}\right]\,S_{i},
\label{Si}
\nonumber
\end{eqnarray}
where $S_{i}$ is defined by the closed paths showed in Fig. \ref{fig2}.  The examples are three possible paths showed in Fig. \ref{fig2}: $S_{2}=S_{ABCHEFA}$, $S_{3}=S_{ABCHGFA}$ and $S_{4}=S_{ADCHEFA}$. These paths are given by
$
S_{2}=\left(\frac{E_{y2}}{\sqrt{\lambda_{2}}}-\frac{E_{y1}}{\sqrt{\lambda_{1}}}\right)\left(\frac{1}{\sqrt{B_{2}}}-\frac{1}{\sqrt{B_{1}}}\right)\nonumber -\left(\frac{E_{y1}}{\sqrt{B_{2}}}-\frac{E_{y2}}{\sqrt{B_{1}}}\right)\left(\frac{1}{\sqrt{\lambda_{2}}}-\frac{1}{\sqrt{\lambda_1}}
\right)$ ,$ S_{3}=\left(\frac{E_{y2}}{\sqrt{\lambda_{2}}}-\frac{E_{y1}}{\sqrt{\lambda_{1}}}\right)\left(\frac{1}{\sqrt{B_{2}}}-\frac{1}{\sqrt{B_{1}}}\right)\nonumber -E_{y2}\left(\frac{1}{\sqrt{B_{2}}}-\frac{1}{\sqrt{B_{1}}}\right)\left(\frac{1}{\sqrt{\lambda_{2}}}-\frac{1}{\sqrt{\lambda_{1}}}\right)$ and $
S_{4}=\left(\frac{E_{y1}}{\sqrt{\lambda_{2}}}-\frac{E_{y2}}{\sqrt{\lambda_{1}}}\right)\left(\frac{1}{\sqrt{B_{2}}}-\frac{1}{\sqrt{B_{1}}}\right)\nonumber -E_{y1}\left(\frac{1}{\sqrt{B_{2}}}-\frac{1}{\sqrt{B_{1}}}\right)\left(\frac{1}{\sqrt{\lambda_{2}}}-\frac{1}{\sqrt{\lambda_{1}}}\right).\nonumber
$
The geometric phase $ \Gamma(C_{2})=\hat{\mathcal{P}}e^{i\gamma_{n}^{\left(i\right)}}$ is  obtained from Eq. (\ref{Si}) and is given by 
\begin{eqnarray}\label{nap}
\Gamma(C_{2})=\hat{\mathcal{P}}\exp\left\{i \oint_{C_{2}} 2u\left[\sqrt{n+\sigma\ell+1}\,\,\delta_{k,\ell+\sigma}+\right. \right. \\ \nonumber \left. \left.\sqrt{n+\sigma\ell}\,\,\delta_{k,\ell-\sigma}\right]\,S_{i},\right\},
\end{eqnarray}
 and it is a non-Abelian geometric quantum phase, where $\hat{\mathcal{P}}$ is the path ordering operator \cite{tg1,zr1}. However, we cannot obtain a general expression for the non-Abelian geometric phase $U(C_{2})$ without specifying the path.  Let us first consider a fixed value of $n$  and choice a set $k, \ell $ in order to make a cyclic adiabatic evolution and, thus, choose one of possible closed paths  $ABCHEFA$, $ABCHGFA$ and $ADCHEFA$ as shown in showed in Fig. \ref{fig2}. A example a similar calculation was made in a other system in Ref.\cite{lbf} However, we cannot obtain
a general expression for the non-Abelian geometric phase (\ref{nap}) without specifying the path. In this way,  to obtain some values we can choice a determined $n$ in order to make a cyclic adiabatic evolution and, thus, choose the closed path $ C_{2}$. 
 One should note that, although the non-Abelian geometric phase given in Eq.(\ref{nap}) is determined by the control parameters $\lambda$ and $B$,  the existence of such phase depends on the control parameter $E_{y}'$. If $E_{y1}=E_{y2}$, we can see that the non-Abelian geometric phase (\ref{nap})  is an identity matrix  . Of course, we can obtain other geometric phases in Eq. (\ref{nap}) by choosing other paths.  Another case is if one supposes the component $E_{x}'$ to be non-zero, therefore we must deal with the four control parameters: $E_{x}'$, $E_{y}'$, $\lambda$ and $B$. In this case, the calculation of the geometric quantum phase is very hard, hence, we  do not discuss it here.
\section{conclusions}
By using the formalism adopted by Feldman and Kahn \cite{feld}, we have shown a way of building displaced states from the Landau quantization of a neutral particle possessing an induced electric dipole moment. The procedure  of constructing the displaced states is  based on adding a constant external electric field to the field configuration  giving rise to the Landau quantization for a neutral particle with an induced electric dipole moment. Besides, we have investigated the arising of Abelian and non-Abelian geometric quantum phases associated with these displaced Fock states during some adiabatic cyclic evolutions.   We claim that we have obtained  the displaced Fock state for Landau quantization for Wei {\it et al.}  Hamiltonian~(\ref{1.3}), and studied the  Abelian and non-Abelian Berry quantum phase for first time in literature. It is worth pointing out that the system has degeneracies in the energy levels can have in energy levels may have Berry phases with non-Abelian structures, as in the case of the analogue of the Landau quantization studied here. We emphasize  that different paths of the Hamiltonian produce a geometric phase that in general does not commute among themselves, which allows a way of implementing the holonomic quantum computation~\cite{zr1,pzr1,pc1,pc3}. As we can see from the  section IV, the space of parameters defined by  $\{E_{x},E_{y},B,\lambda\}$ enables us to have a greater number of possible ways to calculate the geometric phase in this space of parameters. In this theoretical study, we follow the  fields configurations inspirited in the original work Wei {\.it et al}\cite{wei,c1lwhw,lin} and we have used the  charge density $\lambda$ as a variable parameter. For a more realistic  experimentally  configuration   these densities can be replaced with a capacitor, where the field  produced by it  induce the dipole moment of the particle(atom), and the deference of potential of this capacitor can be used as variable parameter, as it has been done recently in the experimental study to obtain the He-Mckellar-Wilkens quantum phase, where the  experimental configuration proposed  by Wei-Han-Wei\cite{wei,c1lwhw} was used by Vigu\'e group\cite{lepoutre,lepoutre1,lepoutre2,lepoutre3} for its experimental detection.This richness  of possibilities to obtain non-Abelian geometric phases is important  for investigation of  holonomic quantum computation in this system of induced dipoles. An interesting point of discussion is the possible use of quantum holonomies defined by the non-Abelian geometric quantum phase in studies of the holonomic quantum computation \cite{zr1} since the present  configuration can be investigated in systems of atoms with a large polarizability. 
 \acknowledgments
The authors would like to thank CNPq, CAPES  for financial support.

\end{document}